 \documentclass[final,1p,times]{elsarticle}
 \usepackage{mathrsfs}

\usepackage{amsmath}
\usepackage{graphics}
\usepackage{txfonts}
\usepackage{setspace}
\usepackage{amsfonts,amssymb,graphicx,mathrsfs,epsfig,subfigure}
\usepackage[dvipdfm,colorlinks,linkcolor=blue,citecolor=blue]{hyperref}
\usepackage{multirow}
\usepackage{booktabs}
\usepackage{fancyhdr,times}

\usepackage{lineno}

\biboptions{sort&compress}

\newproof{proof}{Proof}
\newdefinition{remark}{Remark}

\begin{document}
\begin{frontmatter}

\title{\Large \bf Psychological effect can lead to bistability in epidemics \tnoteref{t1}}

 \tnotetext[t1]{This work is supported by NSFC (No.U1604180), Key Scientific and Technological
Research Projects in Henan Province (No.192102310089), Foundation of
Henan Educational Committee (No.19A110009) and Grant of
Bioinformatics Center of Henan University (No.2019YLXKJC02).}
\author[els]{Shaoli Wang \corref{cor1}}
\ead{wslheda@163.com }
\author[els]{Xiyan Bai  }

 \cortext[cor1]{Corresponding author.}

\address[els]{School of Mathematics and Statistics, Henan University, Kaifeng 475001, Henan, PR China }

\begin{abstract}

\begin{spacing}{1.0}

In this paper, we study the psychological effect in a SIS epidemic
model. The basic reproduction number is obtained. However, the
disease free equilibrium is always asymptotically stable, which
doesn't depends on the basic reproduction number. The system has a
saddle-node bifurcation appear and displays bistable behavior, which
is a new phenomenon in epidemic dynamics and different from the
backward bifurcation behavior.

 \end{spacing}

\end{abstract}


\end{frontmatter}

\section{Introduction}

In classic disease transmission model, the incidence rate is
bilinear in the infectious fraction $I$ and the susceptible fraction
$S$. Recently nonlinear incidence functions in epidemic models
attracted much attention
\cite{Capasso,Liu,Liu2,Hethcote2,Ruan1,Ruan2,Li,Huang,Xiao,
Hethcote,Busenberg,Derrick,Hethcote3}.

 Capasso and Serio \cite{Capasso}, Ruan and Wang \cite{Ruan1} show the incidence function $g(I)$ can interpret the ``psychological"
effect: for a very large number of infective individuals the
infection force may decrease as the number of infective individuals
increases, because in the presence of large number of infective the
population may tend to reduce the number of contacts per unit time.
Xiao and Ruan \cite{Xiao} studied an epidemic model with
nonmonotonic incidence rate, which describes the psychological
effect of certain serious diseases on the community when the number
of infectives is getting larger. Lu et al. \cite{Huang} provided a
more reasonable incidence function, which first increases to a
maximum when a new infectious disease emerges or an old infectious
disease reemerges, then decreases due to psychological effect, and
eventually tends to a saturation level due to crowding effect.

In this paper we will discuss the psychological effect in epidemics
in a different way. The general SIS epidemic model takes the
following form
$$\left\{ \begin
{array}{l l} \frac{dS}{dt}=b-dS-k_{1}Sg(I)+\gamma I, \\
\frac{dI}{dt}=k_{1}Sg(I)-(d+\mu+\gamma)I,
  \end {array} \right.  \eqno(1.1)$$
where $b$ is  natural birth rate, $d$ is  natural decay rate,
$k_{1}$ is  transmission rate for naive susceptible, $\mu$ is
disease related death rate, $\gamma$ is the  rate of infective
individuals lose immunity and move into susceptible compartment. For
the incidence rate $g(I)$, we have following cases, some are based
on the work of Andrews \cite{Andrews}.

 (I) If $g(I)=I$, then system (1.1) is the classic SIS model;

(II) If we choose $g(I)S$ as following
$$g(I)S=\frac{S}{1+\frac{k_{s}}{I}}=\frac{SI}{k_{s}+I},$$
which is the saturated incidence rate in epidemic models
\cite{Capasso,Hethcote,Busenberg}. Here $k_{s}$ is the saturation
constant of infected population concentration;

(III) If
$$g(I)S=\frac{S}{1+\frac{k_{s}}{I}+\frac{I}{k_{i}}}=\frac{k_{i}SI}{k_{i}k_{s}+k_{i}I+I^{2}},$$
 which
is the nonmonotone incidence rate, where  $k_{i}$ is the inhibition
constant of infected population concentration. The special case is
$$g(I)S=\frac{S}{\frac{k_{s}}{I}+\frac{I}{k_{i}}}=\frac{k_{i}SI}{k_{i}k_{s}+I^{2}}=\frac{\frac{1}{k_{s}}SI}{1+\frac{1}{k_{i}k_{s}}I^{2}},$$
which was studied by Xiao and Ruan \cite{Xiao}.

(IV) If we choose $g(I)S$ as following
$$g(I)S=\frac{SI}{1+\frac{k_{s}}{I}+\frac{I}{k_{i}}}=\frac{k_{i}SI^{2}}{k_{i}k_{s}+k_{i}I+I^{2}},$$
which is the  generalized nonmonotone and saturated incidence rate
\cite{Huang}. Especially, if
$$g(I)S=\frac{SI}{\frac{k_{s}}{I}+\frac{I}{k_{i}}}=\frac{k_{i}SI^{2}}{k_{i}k_{s}+I^{2}}=\frac{\frac{1}{k_{s}}SI^{2}}{1+\frac{1}{k_{i}k_{s}}I^{2}},$$
which was studied by Ruan and Wang \cite{Ruan1}, and Tang et
al.\cite{Ruan2}.

(V) If we choose $g(I)S$ as following
$$g(I)S=\frac{SI^{r}}{1+\frac{k_{s}}{I^{p}}+\frac{I^{p'}}{k_{i}}}=\frac{k_{i}SI^{p+r}}{k_{i}k_{s}+k_{i}I^{p}+I^{q}},$$
which is the general incidence rate. Here $r, p, p'$ and $q=p+p'$
are nonnegative. The special case is
$$g(I)S=\frac{SI^{r}}{\frac{k_{s}}{I^{p}}+\frac{I^{p'}}{k_{i}}}=\frac{k_{i}SI^{p+r}}{k_{i}k_{s}+I^{q}}=\frac{\frac{1}{k_{s}}SI^{p+r}}{1+\frac{1}{k_{i}k_{s}}I^{q}},$$
which was studied by a number of authors
\cite{Liu2,Derrick,Hethcote2,Hethcote3}.

 In this paper, we only consider the case $p=1, q=2, r=1$.  Denote $k_{1}k_{i}=k, k_{i}k_{s}=\alpha, k_{i}=\beta,$ then
system (1.1) can be written by following model

$$\left\{ \begin
 {array}{l l} \frac{dS}{dt}=b-dS-\frac{kSI^{2}}{\alpha+\beta I+ I^{2}}+\gamma I, \\
\frac{dI}{dt}=\frac{kSI^{2}}{\alpha+\beta I+ I^{2}}-(\mu+\gamma+d)I. \\
  \end {array}  \right.
   \eqno(1.2) $$
Here,  $\beta$ is the inhibition psychological effect constant of
infected population and $\alpha, \beta$ are positive.

\section{Equilibria and thresholds}

It can be verified that the nonnegative orthant $\mathbb{R}^{+}_{2}
=\{(S, I): S\geq0, I>0\}$ is positively invariant with respect to
system (1.2) and the model is well posed.

Denote $$ R_{0}=\frac{bk}{\beta d(\mu+\gamma+d)}=\frac{b}{d}\cdot
k_{1}\cdot\frac{1}{\mu+\gamma+d}$$ be the basic reproduction number,
which  determining whether or not the disease dies out in
 classical SIS epidemic models. We also denote
$$R_{c}=R_{0}-\frac{2}{\beta}\sqrt{\alpha(1+\frac{k(\mu+d)}{d(\mu+\gamma+d)})},$$ and $$
R_{cc}=R_{0}+\frac{2}{\beta}\sqrt{\alpha(1+\frac{k(\mu+d)}{d(\mu+\gamma+d)})}.$$
It is easy to see that $R_{c}<R_{0}<R_{cc}$.

(i) System (1.2) always has a disease-free equilibrium
$E_{0}=(\frac{b}{d}, 0)$.

(ii)  To obtain the positive equilibria of system (1.2), we solve
the following equations:

$$\begin
{array}{l l} b-dS-\frac{kSI^{2}}{\alpha+\beta I+ I^{2}}+\gamma I=0, \\
\frac{kSI}{\alpha+\beta I+ I^{2}}-(\mu+\gamma+d)=0. \\
  \end {array}
   \eqno(2.1) $$
Solving the first equation of (2.1), we have
$$S=\frac{b-(\mu+d)I}{d},$$
substituting which into the second equation of (2.1) yields

$$AI^{2}+BI+\alpha=0, \eqno(2.2) $$
where $$A=1+\frac{k(\mu+d)}{d(\mu+\gamma+d)},$$
$$B=\beta(1-R_{0}).$$
Denote $\Delta=B^{2}-4A\alpha.$ If $\Delta>0$, then $R_{cc}<1$ or
$R_{c}>1.$ If $B<0$, then $R_{0}>1$.
 When $R_{c}>1$, equation (2.2) has two positive
roots:
$$ I_{\pm}^{*}=\frac{-B\pm\sqrt{\Delta}}{2A}.$$

\noindent{\bf Theorem 2.1}\hspace{0.1cm} (i) System (1.2) always has
a disease-free equilibrium $E_{0};$

(ii) If $R_{c}>1$, system (1.2) also has two positive equilibria
$E_{+}^{*} (S_{+}^{*}, I_{+}^{*}), E_{-}^{*} (S_{-}^{*},
I_{-}^{*}),$ where

$$S_{+}^{*}=\frac{b-(\mu+d)I_{+}^{*}}{d}, I_{+}^{*}=\frac{-B+\sqrt{\Delta}}{2A}, $$
$$S_{-}^{*}=\frac{b-(\mu+d)I_{-}^{*}}{d}, I_{-}^{*}=\frac{-B-\sqrt{\Delta}}{2A}. $$

The existence of positive equilibria are summarized in Table $1$.

\begin{table*}[ht]
\caption{The existence of the positive equilibria  of system (1.2)}
\begin{center}
\begin{tabular}{|l|ll|l|}
\hline   & $R_{c}<1$  &$R_{c}>1$
\\\hline
$E_{0}$                   &exist     &exist     \\
 $E_{+}^{*}$      &  --- & exist     \\
 $E_{-}^{*}$        &  --- &  exist    \\
\hline
\end{tabular}
\end{center}
\end{table*}

\section{Stability analysis}

Let $\widetilde{E}$ be any arbitrary equilibrium of system (1.2).
The Jacobian matrix associated with  system (1.2) is

$$\mathscr{J}_{\widetilde{E}}=\left[
\begin{array}{cccc}
 -d-\frac{k\widetilde{I}^{2}}{\alpha+\beta \widetilde{I}+ \widetilde{I}^{2}}     &\gamma-\frac{k\widetilde{S}\widetilde{I}(2\alpha+\beta \widetilde{I})}{(\alpha+\beta \widetilde{I}+ \widetilde{I}^{2})^{2}}       \\
\frac{k\widetilde{I}^{2}}{\alpha+\beta \widetilde{I}+ \widetilde{I}^{2}}        &\frac{k\widetilde{S}\widetilde{I}(2\alpha+\beta \widetilde{I})}{(\alpha+\beta \widetilde{I}+ \widetilde{I}^{2})^{2}} -(\mu+\gamma+d)     \\

\end{array}
\right].$$ The characteristic equation of  system (1.2) at
$\widetilde{E}$ is  $\left|\lambda I-\mathscr{J}_{\widetilde{E}}
\right|=0.$

\textbf{3.1. Stability analysis of  the disease-free equilibrium }

\noindent{\bf Theorem 3.1 }\hspace{0.1cm} The disease-free
equilibrium $E_{0}$ of system (1.2) is always locally asymptotically
stable.

\noindent{\bf Proof.} The characteristic equation of system of (1.2)
at the disease-free equilibrium $E_{0}$ is obtained as
$$(\lambda +d)(\lambda+\mu+\gamma+d)=0.$$
The characteristic polynomial  has two roots $-d$,
$-(\mu+\gamma+d)$. Since the two roots are all negative,  the
disease-free equilibrium $E_{0}$ of system (1.2) is locally
asymptotically stable.\qed

\textbf{3.2. Stability analysis of positive equilibria}

 \noindent{\bf Theorem 3.2}\hspace{0.1cm} If $R_{c}>1$,
$a_{1}>0,$ system (1.2) has two positive equilibria $E_{+}^{*}$ and
$E_{-}^{*}$, where $E_{+}^{*}$ is a locally asymptotically stable
and $E_{-}^{*}$ is  unstable.

\noindent{\bf Proof.} Denote  an arbitrary positive equilibrium of
system (1.2) as $E^{*}$. The characteristic equation of the system
(1.2) at the arbitrary positive equilibrium $E^{*}$ is obtained as

$$\lambda^{2}+a_{1}\lambda+a_{2}=0,$$
where
$$\begin {array}{lll}
a_{1}=\mu+\gamma+2d+\frac{k(I^{*})^{2}}{\alpha+\beta I^{*}+ (I^{*})^{2}}-\frac{(\mu+\gamma+d)(2\alpha+\beta I^{*})}{\alpha+\beta I+(I^{*})^{2}},\\
a_{2}=d(\mu+\gamma+d)+(\mu+\gamma+d)\frac{k(I^{*})^{2}}{\alpha+\beta I^{*}+ (I^{*})^{2}}-d\frac{(\mu+\gamma+d)(2\alpha+\beta I^{*})}{\alpha+\beta I^{*}+ (I^{*})^{2}}-\gamma \frac{k(I^{*})^{2}}{\alpha+\beta I^{*}+(I^{*})^{2}}.\\
\end {array}$$

(i) For equilibrium $E_{+}^{*},$ we have
$$\begin {array}{lll}
d (I_{+}^{*})^{2}+k(I_{+}^{*})^{2}-d\alpha-\gamma\frac{k(I_{+}^{*})^{2}}{(d+\delta)(d+\mu)},\\
=d(1+\frac{k}{d}(1-\frac{\gamma}{\mu+\gamma+d}))(I_{+}^{*})^{2}-d\alpha,\\
=d(1+\frac{k(\mu+d)}{d(\mu+\gamma+d)})\frac{(\beta(R_{0}-1)+\sqrt{\Delta})^{2}}{4A^{2}}-d\alpha,\\
=\frac{d(\beta(R_{0}-1)+\sqrt{\Delta})^{2}}{4A}-d\alpha.\\
\end {array}$$
It follows from
$$\frac{d(\beta(R_{0}-1)+\sqrt{\Delta})^{2}}{4A}-d\alpha=\frac{d\Delta+d\beta (R_{0}-1)\sqrt{\Delta}}{2A}$$  that
$\frac{d(\beta(R_{0}-1)+\sqrt{\Delta})^{2}}{4A}-d\alpha>0$. Then,
$$\begin {array}{lll}
\frac{d(\beta(R_{0}-1)+\sqrt{\Delta})^{2}}{4A}-d\alpha>0,\\
\Leftrightarrow d+\frac{k(I_{+}^{*})^{2}}{\alpha+\beta I_{+}^{*}+(I_{+}^{*})^{2}}-\frac{d(2\alpha+\beta I_{+}^{*})}{\alpha+\beta I_{+}^{*}+ (I_{+}^{*})^{2}}-\frac{\gamma}{\mu+\gamma+d}\cdot \frac{k(I_{+}^{*})^{2}}{\alpha+\beta I_{+}^{*}+(I_{+}^{*})^{2}}>0,\\
\Leftrightarrow a_{2}>0.\\
\end {array}$$
 Clearly, $a_{2}>0,$ and we also have $ a_{1}>0$. By the Routh-Hurartz Criterion, we know that the positive equilibrium $E_{+}^{*}$
is a locally asymptotically stable node.

(ii) For equilibrium $E_{-}^{*},$  we have

$$\begin {array}{lll}
d (I_{-}^{*})^{2}+k(I_{-}^{*})^{2}-d\alpha-\gamma\frac{k(I_{-}^{*})^{2}}{(d+\delta)(d+\mu)},\\
=d(1+\frac{k}{d}(1-\frac{\gamma}{\mu+\gamma+d}))(I_{-}^{*})^{2}-d\alpha,\\
=d(1+\frac{k(\mu+d)}{d(\mu+\gamma+d)})\frac{(\beta(R_{0}-1)-\sqrt{\Delta})^{2}}{4A^{2}}-d\alpha,\\
=\frac{d(\beta(R_{0}-1)-\sqrt{\Delta})^{2}}{4A}-d\alpha,\\
<\frac{d[2\beta^{2}(R_{0}-1)^{2}-8A\alpha-2(\beta^{2}(R_{0}-1)^{2}-4A\alpha)]}{4A}=0.\\
\end {array}$$
Thus, $a_{2}<0$. By the Routh-Hurartz Criterion, we know in this
case the positive equilibrium $E_{-}^{*}$ is an unstable saddle.
\qed

\begin{table*}[ht]
\caption{The stabilities of the equilibria and the behaviors of
system (1.2) .  }
\begin{center}
\begin{tabular}{|l|lll|l|}
\hline    &$E_{0}$ & $E_{+}^{*}$ &  $E_{-}^{*}$  &System (1.2)
\\\hline
$R_{c}<1$    & LAS &---  &  ---     & Converges to $E_{0}$ \\
 $R_{c}>1$      &LAS &LAS &US   & Bistable\\
\hline
\end{tabular}
\end{center}
\end{table*}

\section{Saddle-node bifurcation}

In this section, we  discuss the bifurcation behavior of system
(1.2). The conditions for saddle-node bifurcation are derived. If
$R_{c}=1$, system (1.2) undergoes a saddle-node bifurcation. The
positive equilibrium $E_{+}^{*}$ and $E_{-}^{*}$ collide to each
other and system (1.2) has a unique instantaneous positive
equilibrium $\bar{E}$. Also one of the eigenvalues of the Jacobian
evaluated at the instantaneous positive equilibrium
  $\bar{E}=(\bar{S},\bar{I})$ is zero. Here
$\bar{S}=\frac{b-(\mu+d)\bar{I}}{d},
\bar{I}=\frac{\beta(R_{0}-1)}{2A}.$

 \noindent{\bf Theorem 4.1} If $R_{c}=1$ or $R_{0}=1+\frac{2}{\beta}\sqrt{\alpha(1+\frac{k(\mu+d)}{d(\mu+\gamma+d)})}\triangleq R_{0}^{[sn]}$, system
(1.2) undergoes a saddle-node bifurcation around instantaneous
positive equilibrium $\bar{E}=(\bar{S},\bar{I})$.

\noindent{\bf Proof.}  Let  $R_{0}$ be the bifurcation parameter. We
use the Sotomayor's theorem to prove that system (1.2) undergoes a
saddle-node bifurcation. The Jacobian matrix at the saddle-node must
have a zero eigenvalue and two eigenvalues with negative real parts.
Let $F=(f_{1},f_{2})^{T}$ with

$$\begin {array}{lll}
f_{1}=b-dS-\frac{kSI^{2}}{\alpha+\beta I+ I^{2}}+\gamma I\\
~~~=b-dS-R_{0}\frac{\beta d(\mu+\gamma+d)SI^{2}}{b(\alpha+\beta I+I^{2})}+\delta I,\\
f_{2}=\frac{kSI^{2}}{\alpha+\beta I+ I^{2}}-(\mu+\gamma+d)I\\
~~~=R_{0}\frac{\beta d(\mu+\gamma+d)SI^{2}}{b(\alpha+\beta I+ I^{2})}-(\mu+\gamma+d)I.\\
\end {array}$$

The Jacobian matrix of system  (1.2) at $\bar{E}$ is given by

$$\mathscr{J}_{\bar{E}}=\left[
\begin{array}{cccc}
-d-\frac{k\bar{I}^{2}}{\alpha+\beta \bar{I}+ \bar{I}^{2}}     &\gamma-\frac{(\mu+\gamma+d)(2\alpha+\beta \bar{I})}{\alpha+\beta \bar{I}+\bar{I}^{2}}       \\
\frac{k\bar{I}^{2}}{\alpha +\beta \bar{I}+\bar{I}^{2}}        &\frac{(\mu+\gamma+d)(2\alpha+\beta \bar{I})}{\alpha+\beta \bar{I}+ \bar{I}^{2}}-(\mu+\gamma+d)    \\
\end{array}
\right].$$

The matrix has a simple zero eigenvalue, which requires that
$det(\mathscr{J}_{\bar{E}})=0$ at $R_{0}=R_{0}^{[sn]}$. If $V$ and
$W$ represent eigenvectors corresponding to the eigenvectors of
$\mathscr{J}_{\bar{E}}$ and $\mathscr{J}_{\bar{E}}^{T}$
corresponding to the zero eigenvalue, respectively, then they are
given by

$$V=\left[
\begin{array}{cccc}
v_{1}\\
v_{2}\\
\end{array}
\right]=\left[
\begin{array}{cccc}
-\frac{\mu+d}{d}\\
1  \\
\end{array}
\right],$$

$$W=\left[
\begin{array}{cccc}
w_{1}\\
w_{2}\\
\end{array}
\right]=\left[
\begin{array}{cccc}
1\\
1+\frac{(\mu+2d)(\alpha+\beta \bar{I}+\bar{I}^{2})}{k\bar{I}^{2}+(\mu+\gamma+d)(\alpha- \bar{I}^{2})}\\
\end{array}
\right].$$ Thus we get

$$F_{R_{0}}(\bar{E},R_{0}^{[sn]})=\left[
\begin{array}{cccc}
-\frac{\beta d(\mu+\gamma+d)\bar{S}\bar{I}^{2}}{b(\alpha+\beta \bar{I}+ \bar{I}^{2})}\\
\frac{\beta d(\mu+\gamma+d)\bar{S}\bar{I}^{2}}{b(\alpha+\beta \bar{I}+\bar{I}^{2})}\\
\end{array}
\right],$$

$$D^{2}F(\bar{E},R_{0}^{[sn]})(V,V)=\left[
\begin{array}{cccc}
\frac{(\mu+\gamma+d)(\beta\bar{I}^{2}+4\alpha\bar{I}+\beta\alpha)}{(\alpha+\beta \bar{I}+ \bar{I}^{2})^{2}}+2\frac{k\bar{I}(2\alpha+\beta \bar{I})}{(\alpha+\beta \bar{I}+ \bar{I}^{2})^{2}}\cdot \frac{\mu+d}{d}\\
-\frac{(\mu+\gamma+d)(\beta\bar{I}^{2}+4\alpha\bar{I}+\beta\alpha)}{(\alpha+\beta \bar{I}+ \bar{I}^{2})^{2}}-2\frac{k\bar{I}(2\alpha+\beta \bar{I})}{(\alpha+\beta \bar{I}+ \bar{I}^{2})^{2}}\cdot \frac{\mu+d}{d}\\
\end{array}
\right].$$ Clearly,

$$W^{T}F_{R_{0}}(\bar{E},R_{0}^{[sn]})=\frac{\mu+2d}{k\bar{I}^{2}+(\mu+\gamma+d)(\alpha- \bar{I}^{2})}\cdot\frac{\beta d(\mu+\gamma+d)\bar{S}\bar{I}^{2}}{b} \neq 0,$$

$$\begin {array}{lll}
W^{T}D^{2}F(\bar{E},R_{0}^{[sn]})(V,V)=-(\frac{(\mu+\gamma+d)(\beta\bar{I}^{2}+4\alpha\bar{I}+\beta\alpha)}{\alpha+\beta \bar{I}+ \bar{I}^{2}}+2\frac{k\bar{I}(2\alpha+\beta \bar{I})}{\alpha+\beta \bar{I}+ \bar{I}^{2}}\cdot\frac{\mu+d}{d})\cdot \frac{\mu+2d}{k\bar{I}^{2}+(\mu+\gamma+d)(\alpha- \bar{I}^{2})}\neq 0.\\
\end {array}$$
Therefore, from the Sotomayor's theorem, system (1.2) undergoes a
saddle-node bifurcation around instantaneous positive equilibrium
$\bar{E}=(\bar{S},\bar{I})$ at $R_{0}=R_{0}^{[sn]}$. Hence, we can
conclude that when the parameter a passes from one side of
$R_{0}=R_{0}^{[sn]}$ to the other side, the number of positive
equilibria of system (1.2) changes from zero to two.

\section{Numerical simulations and Discussion }

To verify our analytical results,  we carry out some numerical
simulations. In the following, we fix the parameter values
 as follows\cite{Xiao,G}:
$$\hspace{0.02cm}  b=1,   d=0.12, k=0.2,
 \gamma=0.05, \mu=0.15, \alpha=0.5.
   \eqno(5.1)$$
If we choose $\beta=3$, the thresholds  $R_{0}^{[sn]}\approx 1.73$
and $R_{c}\approx 1.005$. In this case, we have a saddle-node
bifurcation (Figure 1).  When $\beta=2.8, R_{0}= 1.86$,  two
equilibria of the model $E_{+}^{*}$ and $E_{0}$ are stable (Figure
2). If we choose $\beta=3.5$, such that $R_{0}= 1.49$, then we have
only one equilibrium $E_{0}$ which is stable (Figure 3);

\begin{figure}[!h]
\begin{center}
{\rotatebox{0}{\includegraphics[width=0.6 \textwidth,
height=40mm]{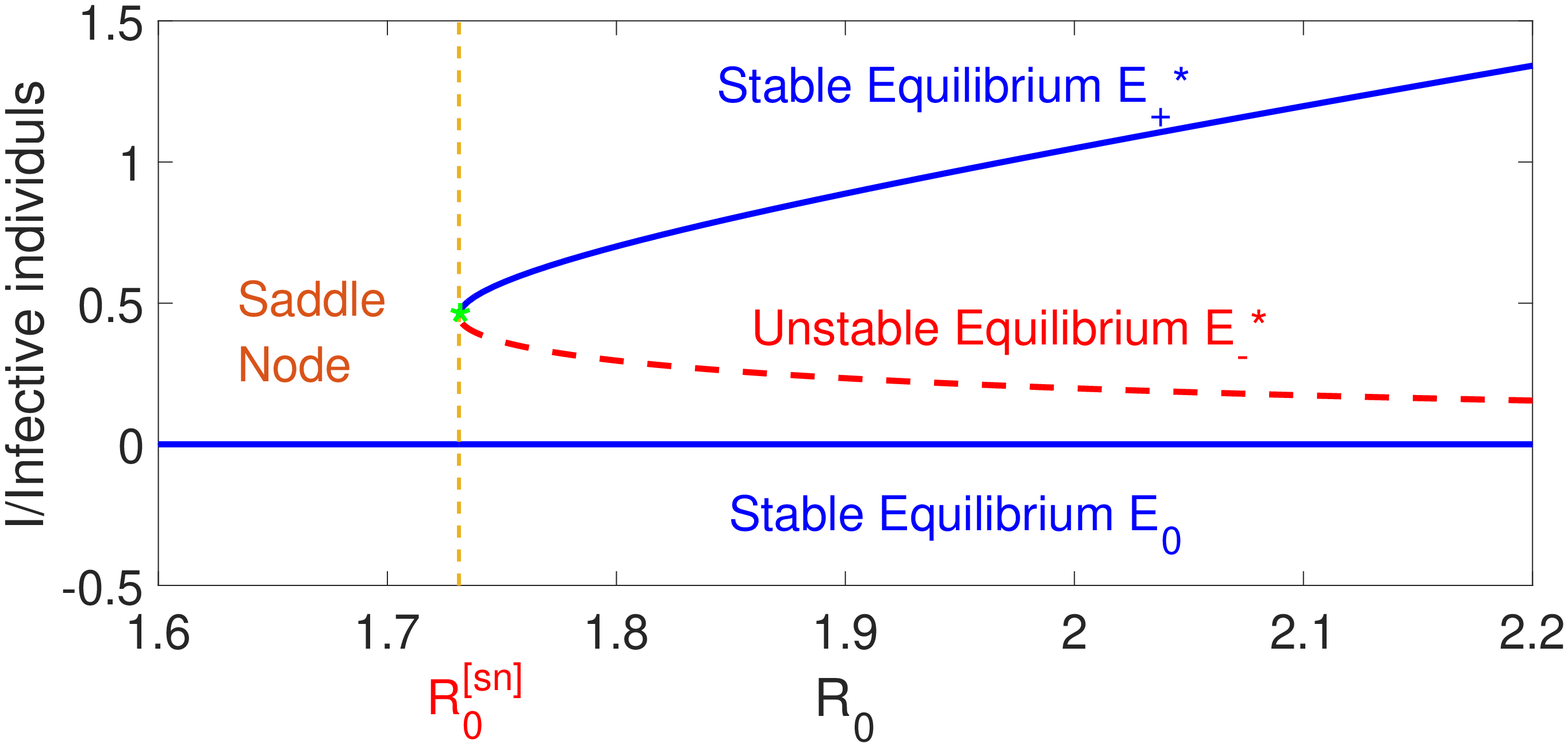}}}
 \caption{
\footnotesize  Bistability and saddle-node bifurcation diagram of
system (1.2). In this case, $R_{0}^{[sn]}\approx 1.73$. The system
displays two stable equilibria $E_{0}$ (the blue solid line at the
bottom) and $E_{+}^{*}$ (the above blue curve), indicating bistable
behaviour. Here, $\bar{E}$ is the saddle point, where the two
equilibria converge and display saddle-node bifurcation. The point
$E_{-}^{*}$ (dashed lines) on the bottom half of the curve is
unstable, and the point $E_{+}^{*}$ (solid line) on the top half of
the curve is stable. Here, $\beta=3$ and other parameter values are
listed in $(5.1)$.}\label{F51}
\end{center}
 \end{figure}

\begin{figure}[!h]
\begin{center}
{\rotatebox{0}{\includegraphics[width=0.40 \textwidth,
height=35mm]{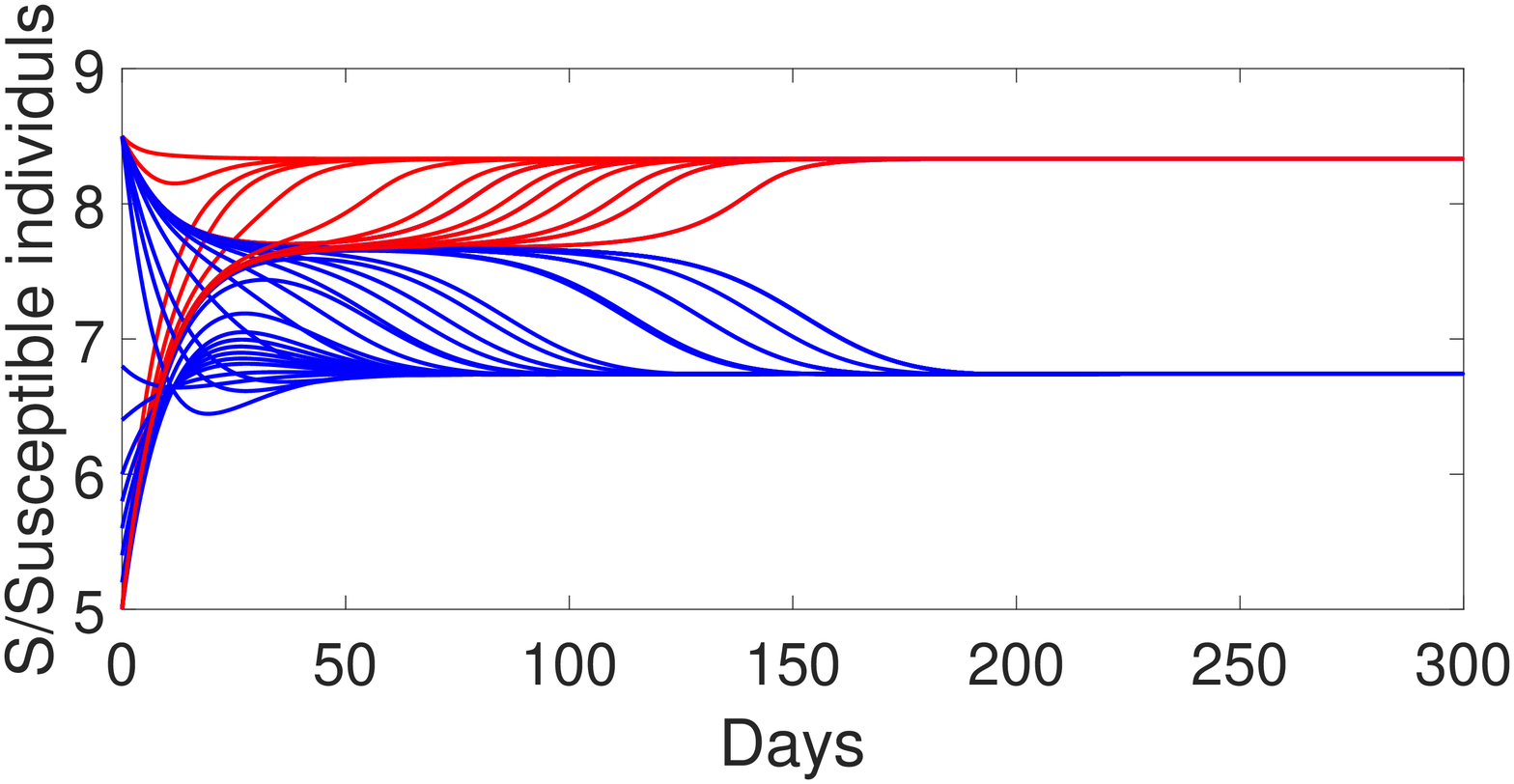}}} {\rotatebox{0}{\includegraphics[width=0.40
\textwidth, height=35mm]{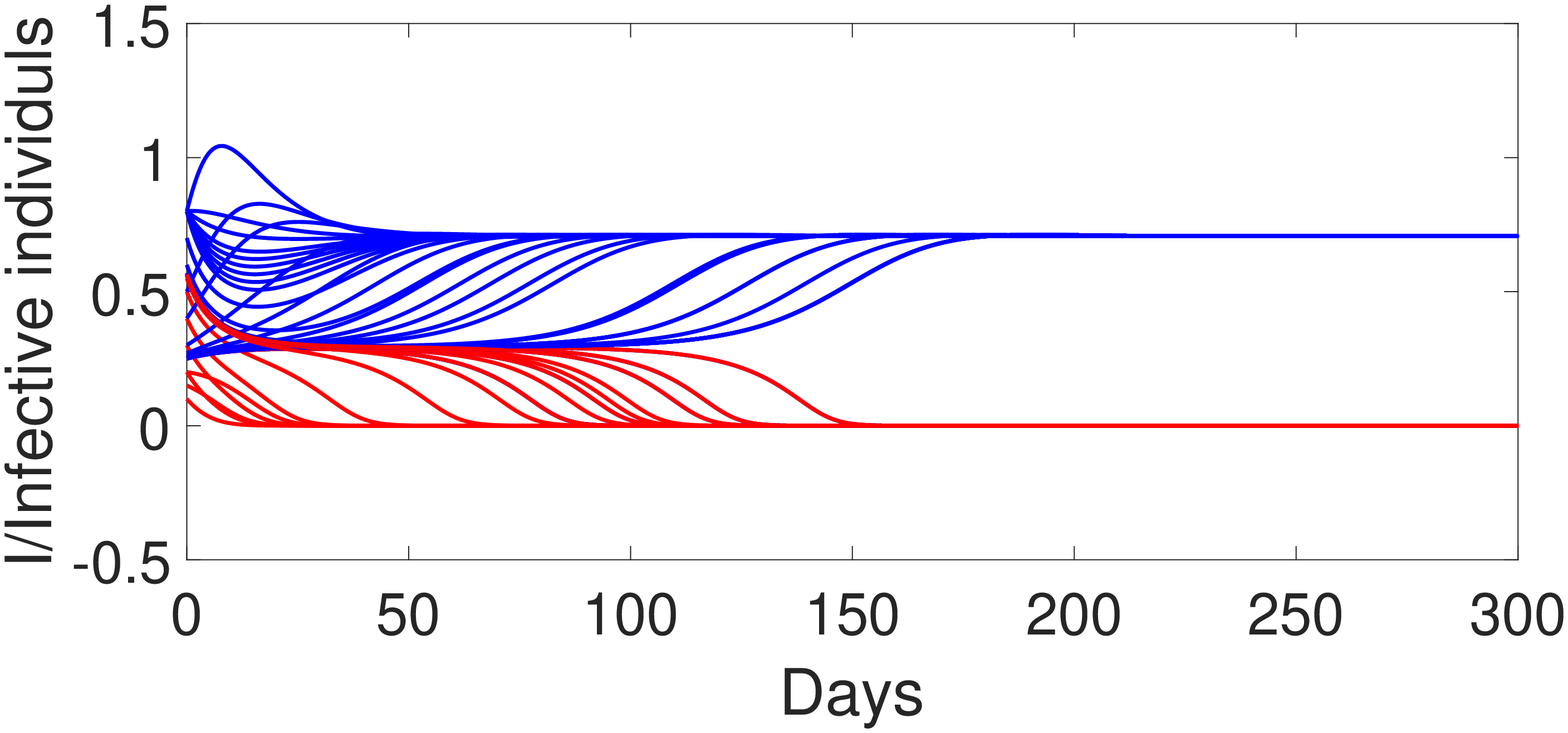}}}
{\rotatebox{0}{\includegraphics[width=0.40 \textwidth,
height=35mm]{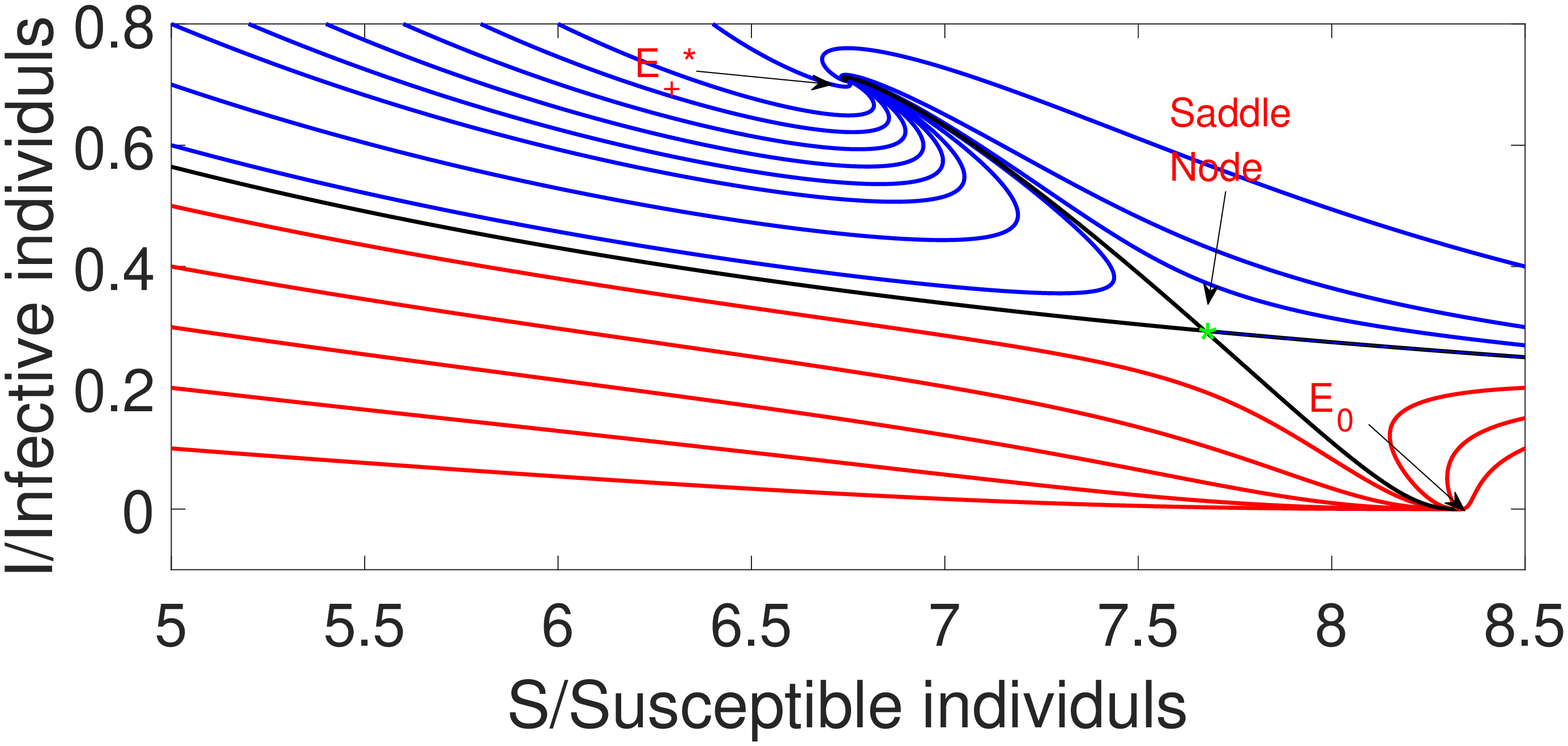}}}
 \caption{
\footnotesize  For $\beta=2.8, R_{0}=1.86$ and other parameter
values listed in (5.1), we can see that in the case of different
initial values, $S$, $I$  converge to either $E_{0}$ or $E_{+}^{*}$.
At this interval, the system display two stable equilibria $E_{0}$
and $E_{+}^{*}$, indicating bistable behavior.}\label{F51}
\end{center}
 \end{figure}

 \begin{figure}[!h]
\begin{center}
{\rotatebox{0}{\includegraphics[width=0.40 \textwidth,
height=35mm]{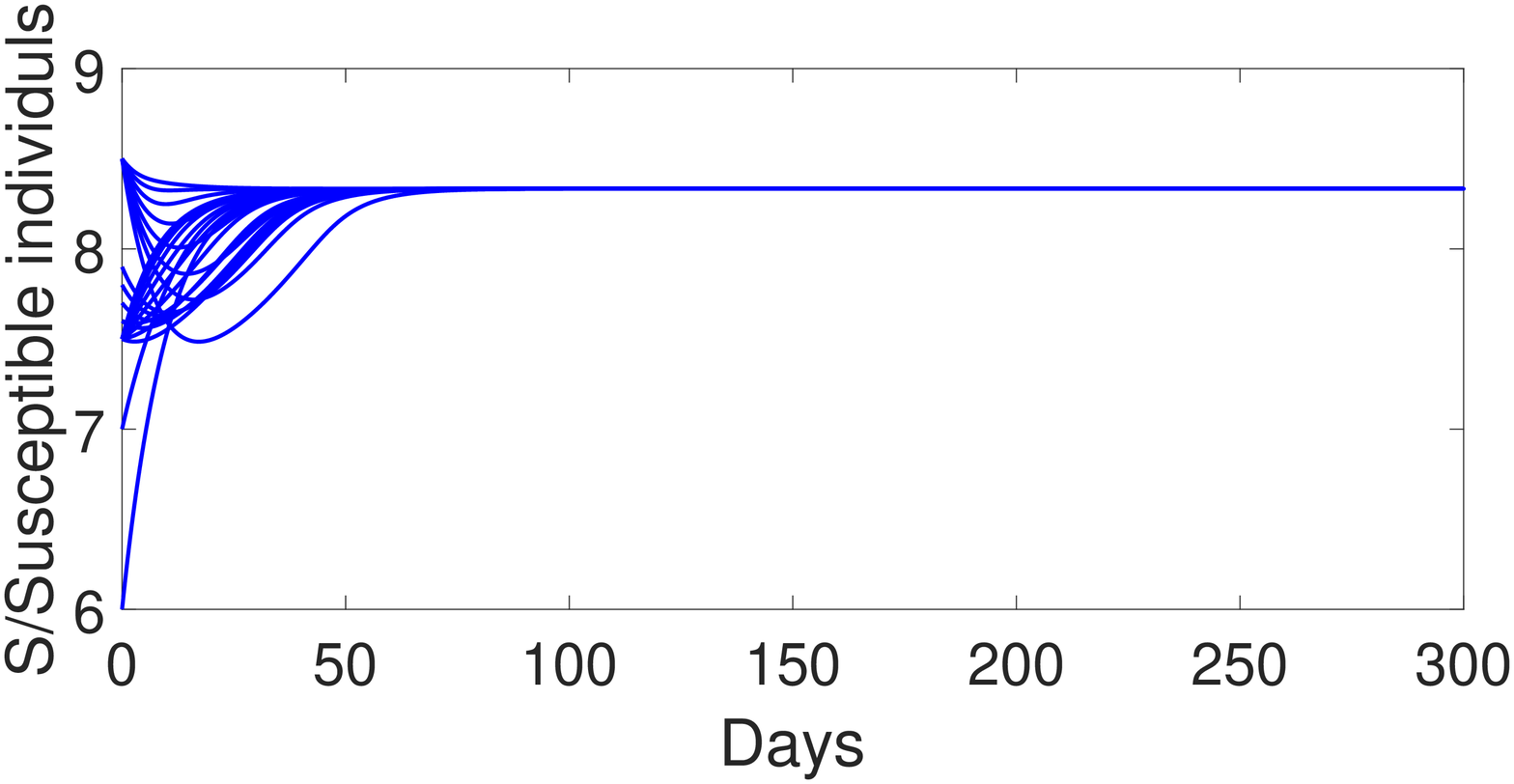}}} {\rotatebox{0}{\includegraphics[width=0.40
\textwidth, height=35mm]{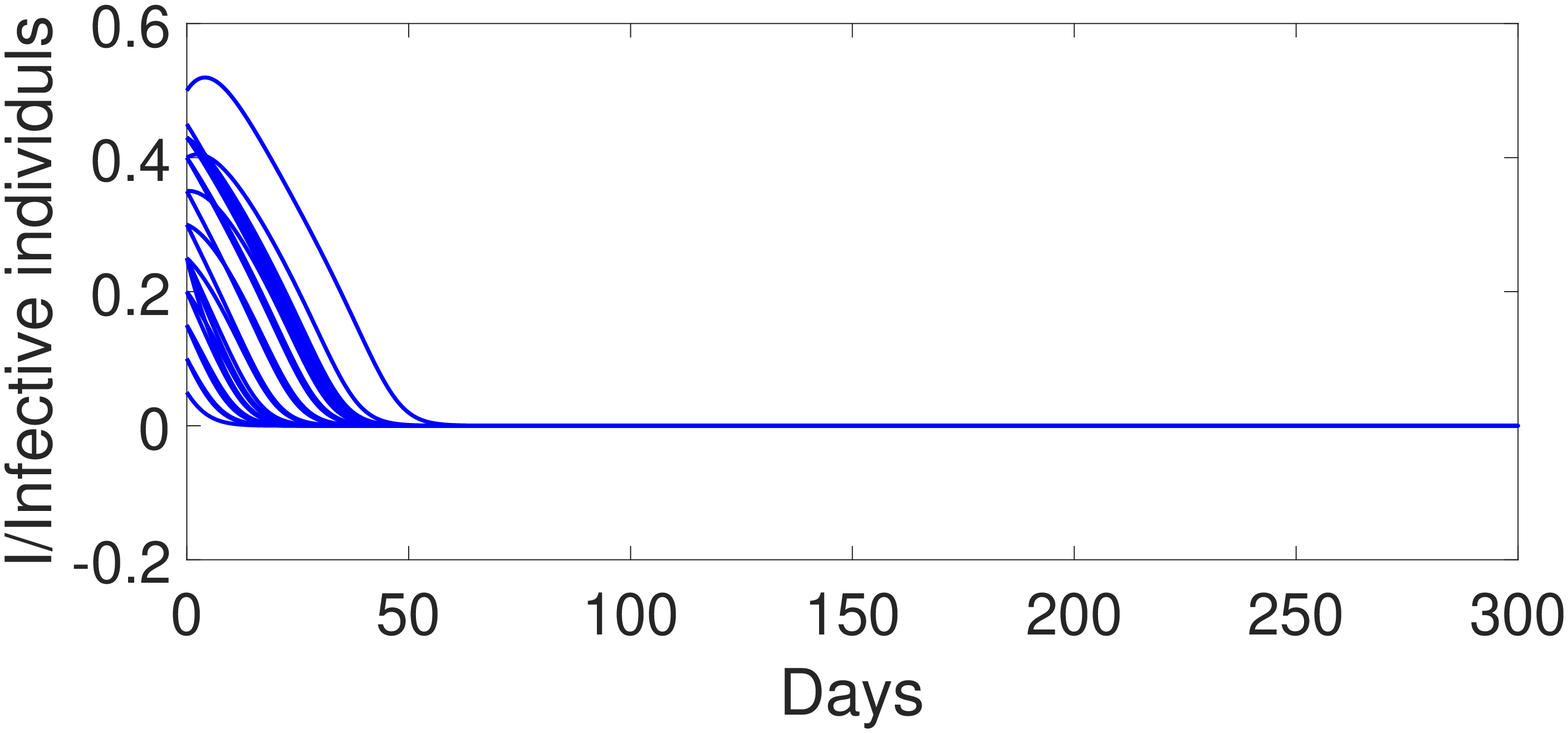}}}
{\rotatebox{0}{\includegraphics[width=0.40 \textwidth,
height=35mm]{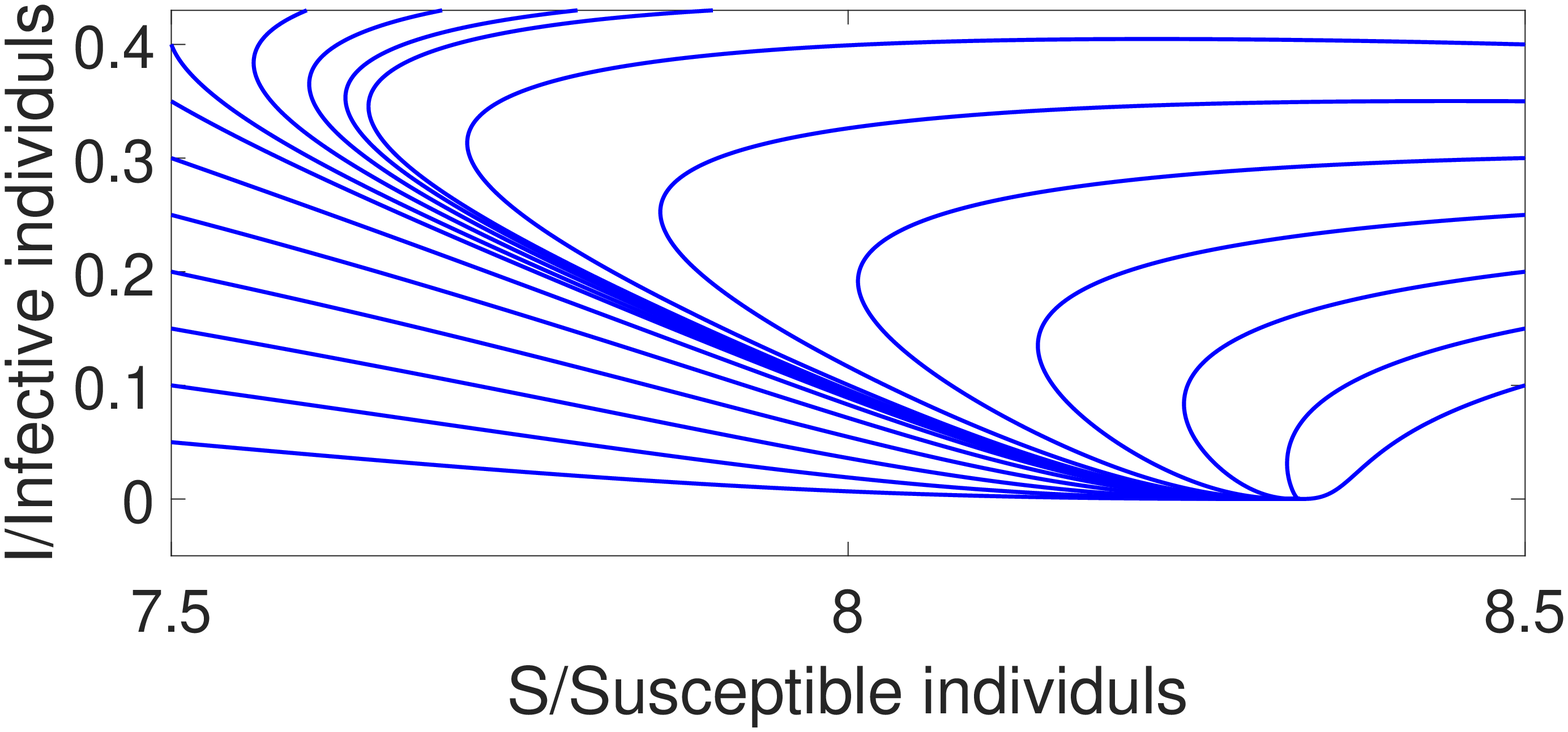}}} \caption{ \footnotesize For $\beta=3.5,
R_{0}=1.49$ and other parameter values listed in (5.1), we can see
that $S$, $I$ converge to $E_{0}$. Here $E_0$  is a locally
asymptotically stable point.}\label{F51}
\end{center}
 \end{figure}

In this paper, we consider a SIS model with psychological effect and
performed mathematical studies. We found that the system displays
bistable behaviors. System (1.2) admits an disease-free equilibrium
$E_{0}$, and two positive equilibria $E_{+}^{*}$ and $E_{-}^{*}$. We
obtain two thresholds,
 the basic reproduction number $R_{0}=\frac{bk_{1}}{d(\mu+\gamma+d)}$ and $R_{c}=\frac{bk}{\beta
d(\mu+\gamma+d)}-\frac{2}{\beta}\sqrt{\alpha(1+\frac{k(\mu+d)}{d(\mu+\gamma+d)})}$.
We find that the system always admits a disease free equilibrium
$E_{0}$ which  is always asymptotically stable, indicating that
there is no infective in the system and all individuals are
susceptible. Thus, if there is no disease, then the uninfected state
will remain stable for a long time. When $R_{c}>1$ , both
$E_{+}^{*}$ and $E_{-}^{*}$ exist, where $E_{+}^{*}$ is locally
asymptotically stable and $E_{-}^{*}$ is unstable, which implies the
coexistence of susceptible, infective individuals. Choosing $R_{0}$
as the branching parameter, our investigation implies that if
$R_{c}=1$ or $R_{0}=R_{0}^{[sn]}$ system (1.2) undergoes a
saddle-node bifurcation. The positive equilibria $E_{+}^{*}$ and
$E_{-}^{*}$ collide to each other and system (1.2) has the unique
instantaneous endemic equilibrium $\bar{E}$. From the branch diagram
in figure 1, we find that when $R_{0}>R_{0}^{[sn]}$, the system has
two stable equilibria $E_{+}^{*}$ and $E_{0}$ appear. The system
displays bistable behavior. When $R_{0}<R_{0}^{[sn]}$, the system
has only one equilibrium point $E_{0}$, suggesting that infectious
diseases will die out eventually.

Castillo-Chavez and Song \cite{Carlos-Castillo-Chavez-Baojun-Song}
proposed the backward bifurcation to illustrate that even if the
basic reproduction number $R_0<1$, disease outbreaks are still
possible. The backward bifurcation indicates that the system
displays bistable behavior when the bifurcation point $R_{c}<R_0<1$.
However, when $R_0 > 1$, the system has only one positive
equilibrium point, which is stable, and the disease-free equilibrium
point is unstable.

In this paper, we investigated a SIS model with psychological
effect. We find that (i) the disease-free equilibrium is always
stable. (ii) When $1<R_0<R_0^{[sn]}$, the model does not have
positive equilibrium point. (iii) When $R_0>R_{0}^{[sn]}$, the
system always display bistability behavior. Our investigation
implies that psychological effect is a kind of self-protection
behavior of human during the outbreak of a disease. Such
self-protection behavior may lead to bistable behavior, i.e., there
may or may not be a disease outbreak.

\end{document}